\documentclass[final,5p,times,twocolumn,10pt]{elsarticle}

\usepackage{amssymb}
\usepackage{amsmath}  
\usepackage{hyperref}  
\usepackage{color}
\usepackage{subfig}

\journal{Nuclear Physics B}

\begin{document}

\begin{frontmatter}



\title{Probing neutrino-nucleus interaction in DUNE and MicroBooNE}%


\author[inst1]{R Lalnuntluanga\corref{cor1}}

\affiliation[inst1]{organization={Department of Physics},
            addressline={Indian Institute of Technology Hyderabad, Kandi}, 
            city={Hyderabad},
            postcode={502284}, 
            state={Telangana},
            country={India}}

\author[inst1]{R K Pradhan}

\author[inst1]{A Giri}

\cortext[cor1]{Corresponding author}



\begin{abstract}
	The neutrino experiments utilize heavy nuclear targets to achieve high statistics neutrino-nucleus interaction event rate, which leads to systematic uncertainties in the oscillation parameters due to the nuclear effects and uncertainties in the cross-section. Understanding the interaction of neutrinos with the nucleus becomes crucial in determining the oscillation parameters with high precision. We investigate the uncertainty in quasi-elastic interaction due to nuclear effects by selecting exactly 1 proton, 0 pions, and any number of neutrons in the final state in DUNE and MicroBooNE detectors, and the effects on the neutrino oscillation in the DUNE detector. The calorimetric method with this selection can be used for accurate neutrino energy reconstruction in the quasi-elastic channel where the nuclear effects are inevitable. 
\end{abstract}




\end{frontmatter}


\section{\label{sec:level1}Introduction}
Neutrino detectors rely on the interaction between neutrinos and the nucleus because neutrinos are not directly detectable. The primary goal of the current and future neutrino experiments is precisely measuring the neutrino oscillation parameters. The neutrino oscillation is dependent on the neutrino energy, and the neutrino flux is not mono-energetic, it spreads in a broad range of energy spectrum due to their production in the decay of produced hadrons. So, an event-by-event neutrino energy reconstruction is required for a better understanding of the uncertainty in neutrino oscillation. Due to high statistics requirements, neutrino experiments used heavy nuclear targets (Argon \cite{ArgoNeuT:2011bms,ArgoNeuT:2014rlj}, Iron \cite{T2K:2014axs,MINERvA:2014rdw}, Water \cite{T2K:2016cbz}, Oxygen \cite{T2K:2014vog}, etc), which arises complications in the reconstruction due to the complexity of the nucleus. Understanding the neutrino-nucleus interactions is crucial to minimize the uncertainties in the neutrino oscillation. 

Most Long-Baseline Neutrino Experiments (LBNE) run in the few GeV energy regions where the quasi-elastic (QE) scattering is one of the most important interaction channels however, there is a significant contribution of 2p-2h interaction which added complications in the cross-section measurement\cite{MiniBooNE:2010bsu,Martini:2009uj}. QE is also a key channel for studying the properties of nucleus apart from the neutrino puzzles. In this work, we focus on the muon neutrino charge current quasi-elastic (CCQE) scattering ($\nu_{\mu}\,n \rightarrow \mu^-\,p$). To understand CCQE scattering, one can refer to impulse approximation (IA) \cite{Benhar:2005dj}, in which interactions occur on the individual nucleons. In the IA, the neutrino interacts with a pair of nucleons inside the nucleus, or a bound nucleon, followed by the final state interaction (FSI) \cite{Dytman:2009zz}. But nucleons inside the nucleus are neither free nor at rest, they are in motion and are bound inside a nuclear potential. As the nucleons are not free particles, their removal energy has to be considered. The scattering with nucleon takes place below the Fermi surface, and excites the nucleon above the Fermi level due to Pauli Blocking \cite{Bodek:2021trq}, where all the nuclear states below the Fermi level are occupied by the valence nucleons, results in knocking out nucleon with momentum above Fermi Momentum. Because of the FSI effect, the hadrons produced in the primary interaction undergo various hadronic scattering such as (in)elastic scattering, pion production, hadron absorption, and charge exchange which results in knocked-out nucleons, and mesons in the final state. This results in misidentification of the QE events as non-QE events. The nucleons interact with each other inside the nucleus through pion exchange, and neutrino interaction with this pair of nucleons leads to multi-nucleon knockout called 2p-2h scattering. Pions produced after FSI in QE could be absorbed inside the nucleus which results in two nucleon knock-out, and can be misidentified as 2p-2h events. Pion production is the major background to the QE scattering events. A CCQE can also produce pions due to FSI and thus cannot be classified as CCQE-like. These backgrounds lead to an uncertainty in the neutrino energy reconstruction, which impacts the oscillation parameters. The effects on the neutrino oscillation studies for experiments such as T2K and MiniBooNE are studied here \cite{Lalakulich:2012hs,Coloma:2013tba}. The non-QE events can also be identified as QE-like because of the nuclear effects. The cross-section of CCQE-like interactions measured in T2K can be explored in Ref. \cite{T2K:2016jor}.

The complication in neutrino interaction due to various nuclear effects is a motivation for Monte Carlo (MC) studies developed with different nuclear models and parameters.
MC event generators are used for predictions and improvement of the experiments. The generators such as GENIE \cite{Andreopoulos:2009rq}, NuWro \cite{Juszczak:2005zs}, GiBUU \cite{Buss:2011mx}, and NEUT \cite{Hayato:2021heg} are extensively used for physics analysis. This work aims to select high-purity CCQE events for DUNE detectors \cite{DUNE:2021tad, DUNE:2023nqi} using the realistic particle thresholds, and its effects on the oscillation parameters are studied using both GENIE and NuWro. Previous studies for LBNE using the GiBUU model can be found in Ref. \cite{Mosel:2013fxa}. The analysis methods for the event selection have been implemented for the MicroBooNE detector \cite{MicroBooNE:2016pwy} using GENIE and NuWro models.

This paper is organized as follows: In section \ref{formalism}, we present the formalism for the neutrino interactions with the nuclear target, and the methods for reconstructing the neutrino energy followed by the neutrino oscillation mechanism. The Monte Carlo event generators are described in section \ref{MC}. The simulation specifications of the DUNE and MicroBooNE detectors are mentioned in section \ref{simulation}. The results from the analysis are discussed in section \ref{results}, followed by the conclusion in section \ref{conclusion}.

\section{Formalism} \label{formalism}
\subsection{Energy Reconstruction}
Consider a charged current neutrino interaction with a nuclear target knocking out \textit{n} nucleons and producing \textit{m} mesons. These produced particles can be used for the neutrino energy reconstruction, using the kinematic and calorimetric methods \cite{Ankowski:2015jya}. The reconstructed energy from the generalized kinematic approach \cite{Ankowski:2015jya} is,
\begin{equation}
	\centering
	E_{\nu} = \frac{2(nM-\epsilon_n)E_l + W^2 - (nM-\epsilon_n)^2 - m_l^2}{2(nM-\epsilon_n-E_l+|k_l| cos \theta)}
\end{equation}
The invariant hadronic mass squared is defined as,
\begin{equation*}
	W^2 = \left(\sum_i E_{p_i'}+ \sum_j E_{h_j'} \right)^2 - \left(\sum_i p_i'+ \sum_j h_j' \right)^2
\end{equation*}
where, $E_{p'}$ and $p'$ ($E_{h'}$ and $h'$) are the energy, and momentum of final state nucleons (mesons) respectively. \textit{M} ($m_l$) is the mass of the final state nucleon (lepton), $E_l$ and $k_l$ are the energy and momentum of the outgoing lepton respectively, and $\theta$ is the scattering angle, $\epsilon_n$ the neutron separation energy.

For a pure charge current quasi-elastic (CCQE) scattering, where the hit nucleon (neutron) is at rest, the kinematic method for reconstructing the neutrino energy \cite{Coloma:2013tba},
\begin{equation} \label{eq:2}
	E_{\nu}=\frac{2(M_n-E_B)E_l -(E_B^2 -2M_nE_B+m_l^2+\Delta M^2)}{2(M_n-E_B-E_l+|k_l| cos \theta)}
\end{equation}
where, $M_n$ is the free neutron rest mass, $\Delta M^2 = M_n^2-M_p^2$, $M_p$ is the rest mass of the proton, and $E_B$ the binding energy. 

On the other hand, the neutrino energy can be reconstructed from the energy of the final state particles using energy and momentum conservation in the calorimetric approach as follows \cite{Ankowski:2015jya},
\begin{equation}
	E_{\nu} = E_{l} + \epsilon_{n} + \sum_{i} ({E_{p'_{i}} - M })+ \sum_{j} {E_{h'_j}}
\end{equation}

Most neutrino experiments rely on the calorimetric method for neutrino energy reconstruction. The comparison of kinematic and calorimetric methods has been studied in Ref. \cite{Ankowski:2015jya}. The calorimetric method is used in this work. For better accuracy in the analysis, rather than assuming constant binding energy value, we considered a distribution for the neutron excitation energy for the nucleus \cite{Furmanski:2016wqo}, for both $^{12}$C and $^{40}$Ar target. A Gaussian distribution of separation energy E with mean $E_{\alpha}$, and deviation $\sigma_{\alpha}$, is considered according to table \ref{tab:1} for $^{12}$C \cite{Frullani:1984nn}, and in table \ref{tab:2} for $^{40}$Ar \cite{Ankowski:2007uy}.

\begin{table}
	\centering
	\begin{tabular}{llll}
		\hline
		Subshell & $E_{\alpha}$ (MeV) & $\sigma_{\alpha}$ (MeV) & No. neutron $n_{\alpha}$ \\ \hline
		1$s_{1/2}$ & 40.8 & 9.1 & 2 \\
		1$p_{3/2}$ & 20.3 & 5   & 4 \\ \hline \hline
	\end{tabular}
	\caption{Neutron Shell Structure in $^{12}$C \cite{Frullani:1984nn}}
	\label{tab:1}
\end{table}

\begin{table}
	\centering
	\begin{tabular}{llll}
		\hline
		Subshell & $E_{\alpha}$ (MeV) & $\sigma_{\alpha}$ (MeV) & No. neutron $n_{\alpha}$  \\ \hline
		1$s_{1/2}$ & 62         & 6.25  & 2         \\
		1$p_{3/2}$    & 40         & 3.75  & 4         \\
		1$p_{1/2}$    & 35         & 3.75  & 2         \\
		1$d_{5/2}$    & 18         & 1.25  & 6         \\
		2$s_{1/2}$    & 13.15      & 1     & 2         \\
		1$d_{3/2}$    & 11.45      & 0.75  & 4         \\
		1$f_{7/2}$    & 5.56       & 0.75  & 2         \\ \hline  \hline
	\end{tabular}
	\caption{Neutron Shell Structure in $^{40}$Ar \cite{Ankowski:2007uy}}
	\label{tab:2}
\end{table}

The probability distribution for the separation energy is given by,
\begin{equation}
	P(E)=\frac{1}{N} \sum_{\alpha} n_{\alpha}G(E-E_{\alpha},\sigma_{\alpha})
\end{equation}
where the sum of neutrons
\begin{equation*}
	N = \sum_{\alpha} n_{\alpha}
\end{equation*}
and $G(E-E_{\alpha},\sigma_{\alpha})$ is the distribution function. While the assumption of a Gaussian distribution for nucleon separation energy is a useful approximation within the scope of our work, it does not reflect the complexity of the physics progress involved as the separation energy extends in the higher energy tail which directly affects the reconstruction of neutrino energy. More details on the consequence of neutron momentum and removal energy distribution on reconstruction methods can be found in Ref. \cite{Benhar:2009wi}.

\subsection{Neutrino Oscillation}
Neutrinos oscillate from one flavor to another while propagating through space as a function of propagating distance, and their energy. Experiments measure oscillation parameters from the neutrino energy distribution at different locations. The event rates measured in the neutrino detector in an accelerator-based experiment depend on the neutrino-nucleus  cross-section as \cite{Nakaya:2020lrq},
\begin{equation}
	N_{\alpha}(E_{\nu}^{reco},L) \approx \int \phi_{\alpha}(E_{\nu}^{true},L) \, \sigma(E_{\nu}^{true}) \, f(E_{\nu}^{true},E_{\nu}^{reco}) \, dE_{\nu}^{true}
\end{equation}

$f(E_{\nu}^{true},E_{\nu}^{reco})$ is the smearing matrix between true and reconstructed neutrino energy that gives uncertainty in the reconstruction. $\sigma(E_{\nu}^{true})$ is the neutrino-nucleus cross-section. $\alpha$ is the neutrino flavor and $\phi_{\alpha}(E_{\nu}^{true},L)$ is the flux at length L i.e. oscillated neutrino flux at the far detector (FD), which is related to near detector (ND) unoscillated flux by
\begin{equation}
	\phi_{\mu}(E_{\nu}^{true},L) \approx P_{\nu_{\mu} \rightarrow \nu_{\mu}}(E_{\nu}^{true},L) \, \phi_{\mu}(E_{\nu}^{true},0)
\end{equation}
$\phi_{\mu}(E_{\nu}^{true},0)$ is the muon neutrino flux at ND, and $P_{\nu_{\mu} \rightarrow \nu_{\mu}}(E_{\nu}^{true},L)$ the survival probability of muon neutrino depends on neutrino energy and baseline length L. For the disappearance channel of LBNE, the muon neutrino survival probability is given by \cite{ParticleDataGroup:2020ssz},

\begin{equation} \label{prob}
	P_{\nu_{\mu}\rightarrow \nu_{\mu}}(E_{\nu},L) \approx 1 - sin^22\theta_{\mu\mu}\,sin^2 \frac{\Delta m^2_{\mu\mu}L}{4E_{\nu}}
\end{equation}
with \begin{equation*}
	sin^2\theta_{\mu\mu}=cos^2\theta_{13}\, sin^2\theta_{23}
\end{equation*}
\begin{equation*}
	\begin{split}
		\Delta m^2_{\mu\mu} = & sin^2\theta_{12} \Delta m_{31}^2 + cos^2\theta_{12}\Delta m_{32}^2 \\
		&+ cos\delta_{CP}\, sin\theta_{13}\,sin2\theta_{12} \,tan\theta_{23}\Delta m_{21}^2
	\end{split}
\end{equation*}

The oscillation parameters are mass squared difference $\Delta m^2$, mixing angle $\theta$, and CP violating phase $\delta_{CP}$. For this study, the values of the oscillation parameters are taken from NuFIT v5.3 \footnote{\href{http://www.nu-fit.org/}{NuFIT 5.3 (2024), www.nu-fit.org.}} \cite{Esteban:2020cvm}. The baseline length for DUNE is $L \approx $ 1300 km.

\begin{figure*}[!htp]
	\centering
	\includegraphics[width=18.5cm,height=7cm]{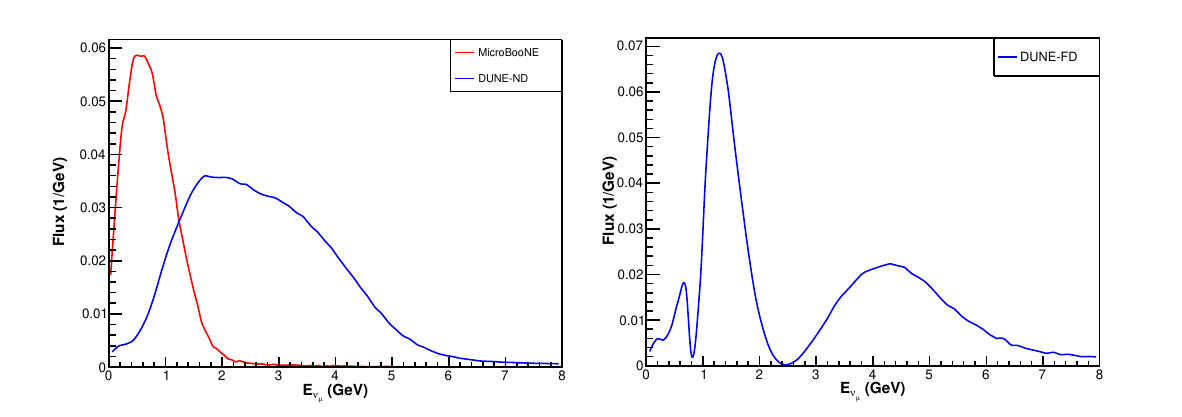}
	\caption{Unoscillated muon neutrino flux for DUNE Near detector and MicroBooNE (left) and oscillated flux for DUNE Far detector (right).}
	\label{fig:1}
\end{figure*}
\section{Monte Carlo Event Generator} \label{MC}
In this work, two MC generators, GENIE and NuWro are used for simulating the neutrino-nucleus cross-section and interactions. GENIE is an experiment-focused event generator widely used by accelerator experiments such as T2K \cite{T2K:2019bcf}, NO$\nu$A \cite{NOvA:2019cyt}, MicroBooNE \cite{MicroBooNE:2016pwy}, MINER$\nu$A \cite{MINERvA:2014rdw}, and MINOS \cite{MINOS:2007ixr}. It satisfies the needs of current oscillation experiments. On the other hand, NuWro is a theory-focused event generator and it is developed to understand the impact of various theoretical frameworks of neutrino interactions on the observables.
The Fermi Gas Model \cite{Bodek:1981wr} is implemented to define the nuclear environment considering nucleon-nucleon correlation effects \cite{Bodek:1980ar} in GENIE, and the spectral function \cite{Benhar:1989aw} in NuWro.

The default model for QE scattering in GENIE is the Llewellyn-Smith (LS) model \cite{LlewellynSmith:1971uhs}. The new models for QE scattering and MEC/2p-2h processes in GENIE are the Valencia QE model \cite{Nieves:2004wx,Gran:2013kda} and SuSAv2 \cite{Dolan:2019bxf,Megias:2016fjk} respectively. The Valencia model is based on the Local Fermi Gas model with Coulomb correction effects, and Random Phase Approximation (RPA). In contrast, the SuSAv2 is based on the Relativistic Mean Field (RMF) theory. In NuWro, the QE interactions are described by the LS model with options for vector and dipole axial vector form factors. GENIE implements the axial mass $M_A$ extending from 0.99 to 1.2 GeV/$c^2$, whereas NuWro confines this parameter in a range of 0.94-1.03 GeV/$c^2$. In this work, the value of the CCQE axial mass is considered as $0.96\, GeV/c^2$ in both GENIE and NuWro, and the vector form factors used are BBA07 \cite{Bodek:2007ym} and BBA05 \cite{Bradford:2006yz} respectively. GENIE has 4 FSI models: hA, hN, INCL++, and Geant4 \cite{geniev3}. The hA model provides an effective description of hadron-nucleus data widely whereas the hN, INCL++, and Geant4 models feature enhanced nuclear medium corrections and incorporate low-energy hadron kinematics.
In NuWro, the FSI is described by the cascade model based on the algorithm by Metropolis \textit{et al.} \cite{Metropolis:1958wvo}. Nuwro considers a semi-classical approach for hadron propagation in the cascade. The hN model in GENIE is similar to the cascade model in NuWro. Empirical MEC model \cite{Katori:2013eoa} is also available for MEC/2p-2h scattering in GENIE, and transverse enhancement model \cite{Bodek:2011ps} and Marteau-Martini model \cite{Martini:2009uj} are also available in NuWro for 2p-2h scattering.

The baryonic resonances are modeled using the Berger-Sehgal (BS) model \cite{Berger:2007rq} in GENIE. NuWro uses the Rein-Sehgal (RS) model \cite{Rein:1980wg} for $\Delta$(1232) resonance, and the Adler-Rarita-Schwinger model \cite{Hernandez:2008qx} for higher resonances. The BS model is similar to the RS model but includes the lepton mass effect. Other available resonance models in GENIE are Kuzmin-Lyubushkin-Naumov (KLN) \cite{Kuzmin:2003ji} and the RS model. 

In the inelastic region (DIS), both the generators use the Bodek-Yang Model \cite{Bodek:2002vp} for the cross-section calculation, and utilize PYTHIA \cite{Sjostrand:2006za} to produce hadronic final states with different values of PYTHIA parameters, such as invariant hadronic mass $W$. GENIE employs it to $W$ = 1.8 or 2.0 GeV with KNO scaling \cite{Koba:1972ng}, while 1.6 GeV for NuWro. The invariant hadronic mass threshold used in this work is $W$ = 1.9 GeV in GENIE, and 1.6 GeV in NuWro.

\section{Simulation Details} \label{simulation}
In this paper, nuclear effects on energy reconstruction are analyzed for the DUNE near and far detector, and the MicroBooNE detector. DUNE mainly uses Argon as the target material. The proposed DUNE near detector HpgTPC \cite{Duffy:2019egx} consists of Argon and Methane (CH$_4$), and liquid Argon as the target material in the far detector. MicroBooNE also uses liquid Argon targets. CC $\nu_{\mu}$ interactions are simulated for the Argon target using the DUNE flux \cite{DUNE:2021cuw}. The DUNE ND flux peaks around 2 GeV, and the FD flux is calculated using the Eq. \ref{prob} taking all the oscillation parameters from NuFIT v5.3 shown in Fig. \ref{fig:1}. The simulation is performed for MicroBooNE using the Argon and Carbon targets with its $\nu_{\mu}$ flux \cite{Fleming:2012gvl} as shown in Fig. \ref{fig:1}. The GENIE v3.01.00 and NuWro-21.09.2 are used for the simulation. We consider realistic simulations and include the Quasi-elastic (QE), Meson exchange (MEC/2p-2h), Resonance (RES), and Deep Inelastic Scattering (DIS) in neutrino mode. In the RES channel, we consider only the first resonance region, $\Delta$ or $P_{33}(1232)$ as it contributes most to the pion production which is the major background for our analysis. For the simulation, we used the GENIE tune G18\_10a \_02\_11a and considered the same models and parameters mentioned in the tune for neutrino interactions in NuWro for consistent results. The models used for neutrino interactions are the Valencia model for QE, the Berger Sehgal model for resonance, and the KNO-tuned Quark-Parton model (PYTHIA) for DIS. For FSI, the hA model is employed in GENIE, and the Oset model in NuWro. To study the FSI effect, we have also simulated the interaction events by switching ON/OFF the FSI in both MC generators. The CCQE axis mass ($M_A$) is considered 0.96 GeV/$c^2$ and the vector mass is 0.84 GeV/$c^2$.

Ideally, a CCQE event can be identified as 1 muon, 1 proton, and 0 pions in the final state. But the pion produced after FSI can be absorbed inside the nucleus knocking out multiple nucleons. We consider a selection of 1muon, 1 proton, 0 pions, and any number of neutrons in the final state \cite{Mosel:2013fxa} for the CCQE interaction. The unobserved neutrons in the detector are a consequence of the nuclear effects. Due to limitations in the sensitivity of detectors, all the produced particles cannot be detected. 
We have applied selection cuts to remove the particles below the detection threshold. The detection thresholds for the DUNE ND are, the total energy cut for muon is 226 MeV \cite{DUNE:2021tad} ($\approx$ 200 MeV momentum), and a minimum kinetic energy threshold of 3 MeV for proton \cite{Hamacher-Baumann:2020ogq}. The neutrons in the detector have a threshold of kinetic energy from 50MeV to 700 MeV \cite{DUNE:2021tad}. In the case of FD, muons have a detector threshold of 30 MeV kinetic energy, and 50 MeV for protons and neutrons \cite{DUNE:2016rla}. In the MicroBooNE detector, the leading proton has a threshold momentum of 300 MeV/c ($\approx$ 47 MeV kinetic energy ), and the momentum threshold for muon is 100 MeV/c \cite{Fleming:2012gvl}. The selection (1 proton + 0 pions + X neutrons) with the corresponding detector thresholds for the DUNE and MicroBooNE detectors are applied to reconstruct the neutrino energy for the CCQE.
\begin{figure*}[hpt]
	\centering
	\includegraphics[width=18cm,height=6.3cm]{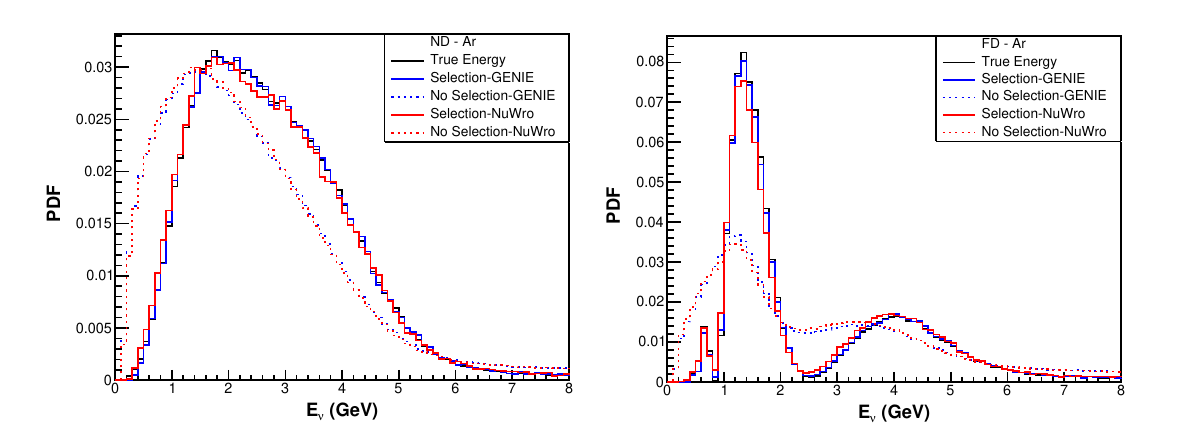}
	\caption{The probability density function (PDF) of reconstructed neutrino energy for Argon in DUNE near (left) and far (right) detector with selection (solid) and without the selection (dotted) using GENIE (blue) and NuWro (red).}
	\label{fig:2}
\end{figure*}

\begin{figure*}[hpt]
	\centering
	\includegraphics[width=18cm,height=6.3cm]{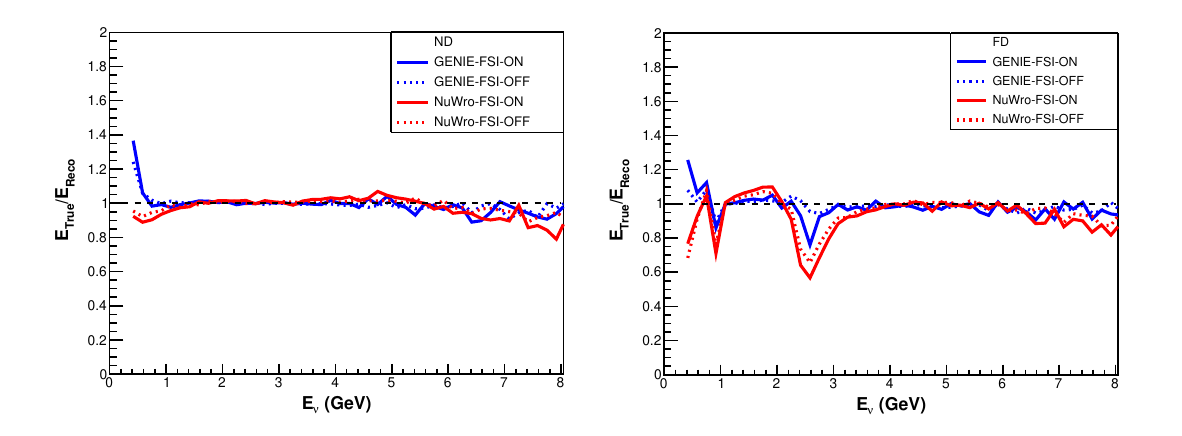}
	\caption{The ratio of true and reconstructed energy PDF as a function of neutrino energy for Argon in DUNE near (left) and far (right) detectors with FSI (solid) and without FSI (dotted) using GENIE (blue) and NuWro (red).}
	\label{fig:3}
\end{figure*}

\begin{figure*}[hpt]
	\centering
	\includegraphics[width=18cm,height=6.3cm]{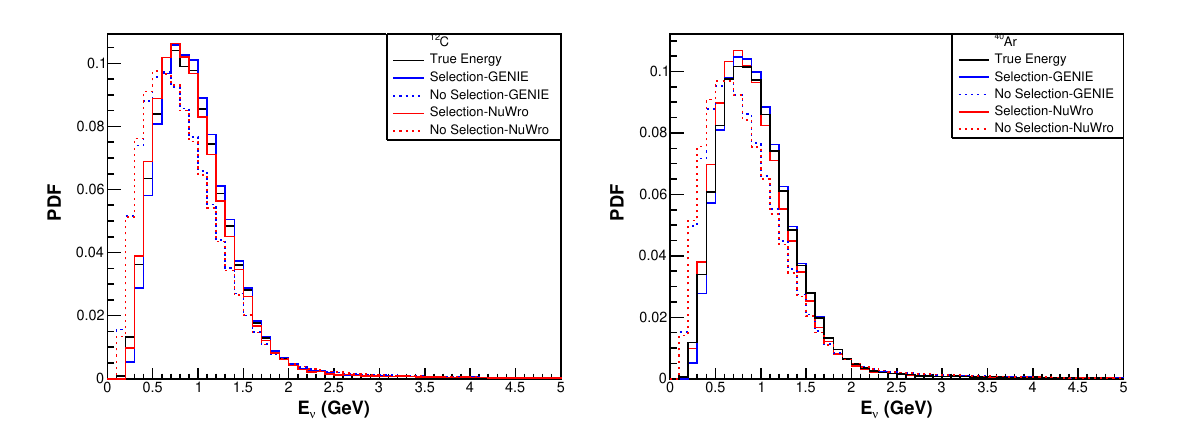}
	\caption{The PDF of reconstructed neutrino energy for Carbon (left) and Argon (right) in MicroBooNE detector with the selection (solid) and without selection (dotted) using GENIE (blue) and NuWro (red).}
	\label{fig:mbenergy}
\end{figure*}

\begin{figure*}[hpt]
	\centering
	\includegraphics[width=18cm,height=6.3cm]{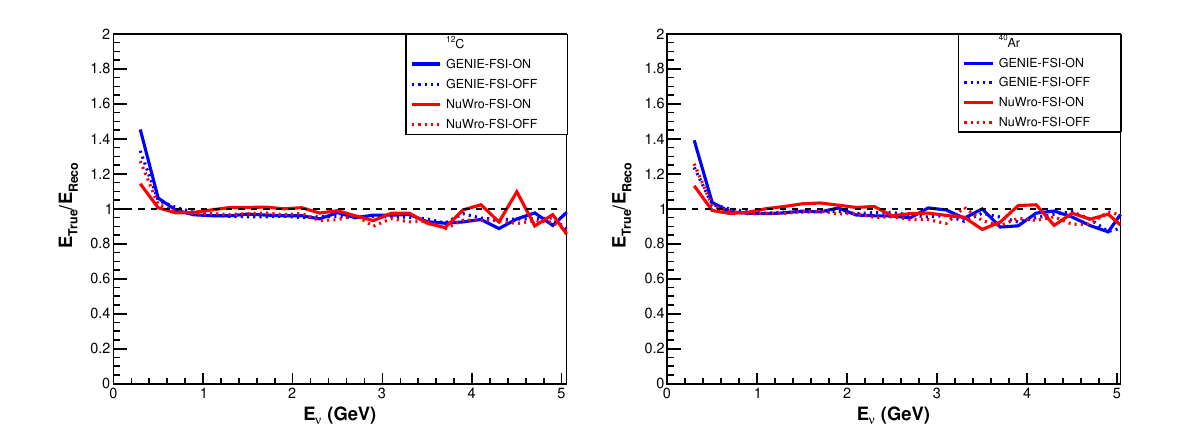}
	\caption{The PDF ratio of true and reconstructed energy as a function of neutrino energy for Carbon (left) and Argon (right) in MicroBooNE detectors with FSI (solid) and without FSI (dotted) using GENIE (blue) and NuWro (red).}
	\label{fig:mbratio}
\end{figure*}

\section{Results} \label{results}
The event distributions of reconstructed neutrino energy for Argon in both DUNE ND and FD are shown in Fig. \ref{fig:2}. For the ND, the reconstructed energy using the calorimetric method without any selection(dotted) is shifted by $\approx$ 0.4 GeV towards lower energies, compared to the true energy distribution (black). The distribution for FD after oscillation is given on the right panel (Fig. \ref{fig:2}). It can be observed clearly that without the selection, the distribution is distorted with flattening of the minima around 2.4 GeV. The event rate is significantly low at the maxima at 1.4 GeV. When the selection (1 proton + 0 pions + X neutrons) is considered along with the DUNE ND and FD detector cuts, the reconstructed energy distribution (solid) agrees with the true energy for both ND and FD. The shift of reconstructed energy is reduced to less than 100 MeV compared to true neutrino energy. The results using GENIE are represented as blue curves. Similarly, NuWro (red) results show the similar effects as GENIE. For evaluating the uncertainty in the reconstruction, we consider the event ratio distribution of true energy with the reconstructed energy shown in Fig.\ref{fig:3}. To further quantify the effect of FSI, we consider the ratio with FSI and without FSI. For the ND, the ratio is close to 1 but in the lower energy region ($<$ 1 GeV), the reconstructed energy differs from the true energy due to the nuclear effects. In case of the far detector, the reconstructed energy disagrees with the true energy in the maxima and minima regions as there is a downward shift in the reconstructed events. When the FSI is switched off, it can be seen that the ratio is closer to unity in the lower energy range, which indicates that the effect of the FSI contributes to the uncertainty in the reconstruction. Even without FSI, the ratio deviates from unity in the lower energy region which could be due be the contribution from other nuclear effects, and 
uncertainty in selection efficiency. Both the generators GENIE and NuWro show this deviation from unity in the lower energy range. The event distributions of reconstructed neutrino energy for Carbon and Argon in the MicroBooNE detector are given in Fig. \ref{fig:mbenergy}. There is a shift of $\approx$ 0.2 GeV in the reconstructed energy without the selection using the calorimetric method. However, the reconstructed energy agrees with true energy when the selection along with detector thresholds are considered from both generators. From the ratio of true neutrino energy to reconstructed energy in Fig. \ref{fig:mbratio}, the deviation from unity can be seen in the lower energy range ($< 0.5$ GeV) even without the FSI effect. The reconstructed neutrino energy for DUNE and MicroBooNE are different according to their neutrino flux, and both the generators show similar results indicating the selection criteria can be considered to identify the CCQE interaction significantly, and the uncertainty could be due to the impact of nuclear effects.
\begin{figure}[h]
	\centering
	\includegraphics[width=9cm,height=7.5cm]{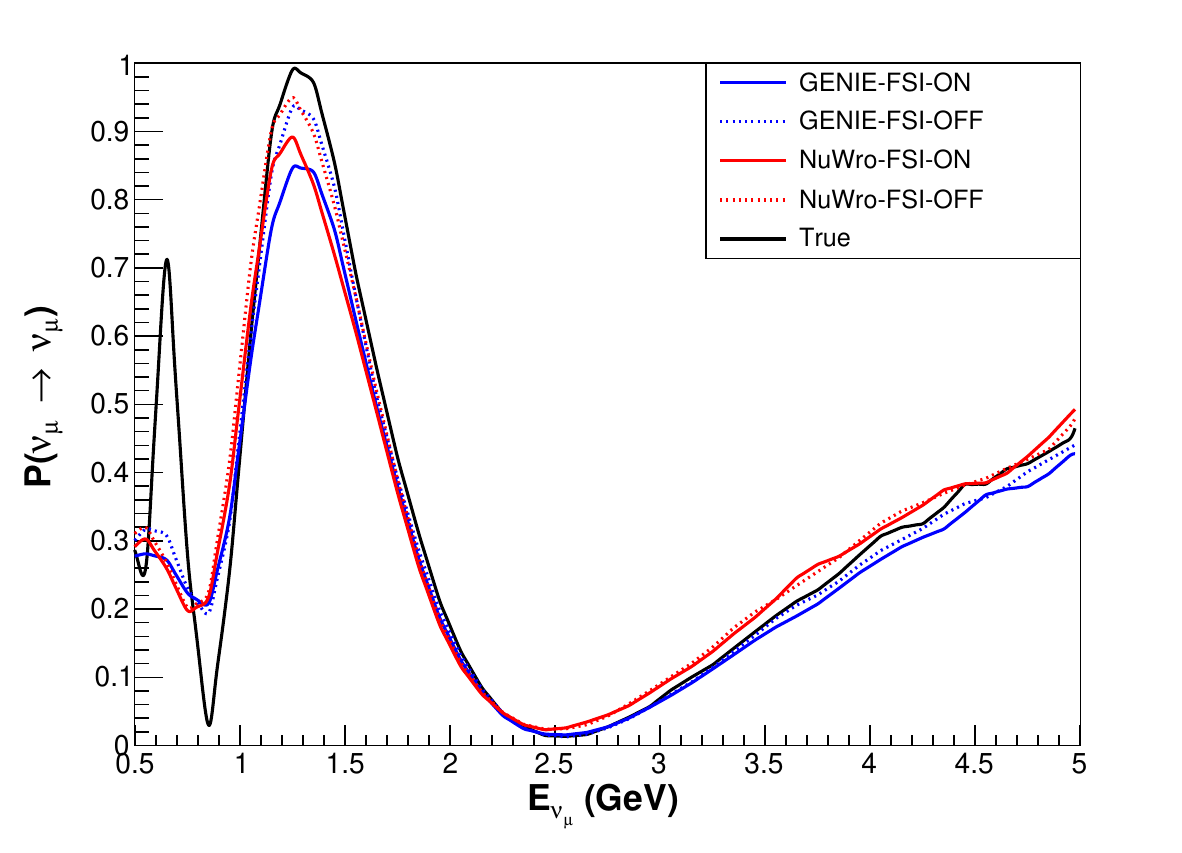}
	\caption{The muon neutrino survival probability in DUNE as a function of true energy without any selection (black) and the reconstructed energy for QE with FSI (solid) and without FSI (dotted) using GENIE (blue) and NuWro (red) considering the selection with detector cuts. }
	\label{fig:4}
\end{figure}

The FSI effects on the neutrino oscillation probability using the selection (1 proton + 0 pions + X neutrons) are studied. The survival probability of muon neutrinos is shown in Fig. \ref{fig:4}. Assuming the experimental analysis, the survival probability is calculated by taking the ratio of oscillated (FD) and unoscillated (ND) event distributions, considering the detector efficiency and energy resolution. In Fig. \ref{fig:4}, the black curve shows the survival probability as a function of true energy, and the blue curves represent the survival probability as a function of the reconstructed energy from GENIE by considering the selection with detector cuts and the FSI effects, and without the FSI is shown by the dotted curve. The red curves show the results from NuWro. There is a significant influence in the absolute values, both at the first maximum and first minimum, while the positions of maximum and minimum are less affected when we consider the selections. The absolute value of the first maximum drops by $\sim$ 10\% when the selection is considered, which could be the uncertainties in detector efficiency, event selection, and cross-sectional uncertainty. There is a clear effect of FSI at the first oscillation peak and the first oscillation deep. when FSI is off, the peak is slightly towards the true one compared to the case where the FSI is on. This indicates a notable discrepancy in the mixing angle, while the impact on $\Delta m^2$ is comparatively negligible. The survival probability as a function of reconstructed energy is away from the true one in the lower energy region below $\sim$ 1.5 GeV because the QE interaction dominates in this energy range. The same effect has been observed for T2K by Coloma et al.  \cite{Coloma:2013rqa} as well. 

\section{Conclusion} \label{conclusion}

The purity of QE interactions selection for the MicroBooNE detector, and the DUNE near and far detectors and their effect on oscillation parameters has been studied using the Monte Carlo neutrino event generators GENIE and NuWro. The physics potential studies on LBNE have shown that an energy resolution of 100 MeV is required to differentiate between different physics properties \cite{LBNE:2013dhi}. The selection of 1 proton, 0 pions, and X neutrons for CCQE shows a weighty discrepancy between the reconstructed and the true neutrino energy less than 100 MeV, which is the required resolution. This shift of less than 100 MeV is observed in both the ND and FD for both generators. The ratio plots deviating from unity even without FSI indicate the other nuclear effects should be quantified in energy reconstruction. This work shows that the calorimetric method can be used for energy reconstruction with the proper selection criteria. The uncertainty in reconstruction due to the nuclear effects, and detector efficiency notably affects the mixing angle, with comparatively lesser effects on $\Delta m^2$.  
Also, the ratio of true and reconstructed energy distribution in the high-energy regions indicates the viability of this method in higher-energy experiments. 

\section*{Acknowledgements}

R Lalnuntluanga thanked the Council of Scientific \& Industrial Research (file number: 09/1001(0054)/2019-EMR-I) for the financial grant. R K Pradhan acknowledges the DST-INSPIRE grant (2022/IF220293) for financial support. R Lalnuntluanga and A Giri credited the grant support of the Department of Science and Technology (SR/MF/PS-01/2016-IITH/G).


 \bibliographystyle{elsarticle-num}
 \bibliography{cas-refs}





\end{document}